\documentclass[aps,superscriptaddress,amsfonts,amssymb,amsmath,showpacs,nofootinbib, twocolumn, floatfix]{revtex4-1}
\usepackage{graphicx}
\usepackage[squaren]{SIunits}

\newcommand{\Eref}[1]{eq.~\eqref{#1}} 
 
\newcommand{\Fref}[1]{Fig.~\ref{#1}} 
 
\newcommand{\Tref}[1]{Table~\ref{#1}} 

\begin{document}

\title{Effect of a variable cosmological constant\\on black hole quasinormal modes}

\author{Cecilia Chirenti} 
\email{e-mail: cecilia.chirenti@ufabc.edu.br} 

\author{Manuela G. Rodrigues} 
\email{e-mail: manuela.rodrigues@ufabc.edu.br} 
\affiliation{Centro de Matem\'atica, Computa\c c\~ao e Cogni\c c\~ao, UFABC, 09210-170 Santo Andr\'e-SP, Brazil}

\begin{abstract}
Many different cosmological models have been proposed to address the cosmological constant problem and the coincidence problem. We compare here four different models that can be used to describe an effective (time-dependent) cosmological constant $\Lambda(z)$. A numerical analysis of the $\Lambda(z)$ evolution obtained for each model shows that it can be used for distinguishing between all four models. We calculate next the $\omega(\Lambda)$ frequencies for quasinormal modes of gravitational perturbations of Schwarzschild-de Sitter black holes at different redshifts. Considering that the variation of $\Lambda$ happens on cosmological timescales, the combined $\omega(\Lambda(z))$ could be used in principle to track the evolution of the cosmological constant. We quantify the resulting minute frequency shift in the quasinormal mode frequencies and show that the relative frequency shift grows as $M^2$. However, even in a most optimistic scenario with an extremely high mass supermassive black hole there is no prospect for the detection of this effect.
\end{abstract}


\maketitle

\section{Introduction}
\label{sec:intro}

According to recent observational evidence, the universe is spatially flat and is currently undergoing an accelerated expansion phase \cite{Riess1998,Perlmutter1999}. The simplest physical explanation for this result would be the existence of a cosmological constant driving the current acceleration of the universe, but there is a huge difference between the observed value of the cosmological constant and the vacuum energy density derived by quantum field theories (the ``cosmological constant problem''). Another problem with the cosmological constant model is the ``coincidence problem'': why should the accelerated expansion occur exactly now in the history of the universe?

These two problems have motivated many different cosmological models, ranging from scalar field models of dark energy to coupled dark energy models and modified gravity alternatives \cite{Copeland2006}. So many different models, based on different physical assumptions, must be constrained by observations. Usual tests include observational constraints from type Ia supernova data, cosmic microwave background radiation and baryon oscillations. However, the comparison between different models should be done carefully, otherwise the results could be model- or calibration-dependent \cite{Seikel2009}.

An effective cosmological ``constant'' can be derived from dynamical dark energy models, its value varying slowly over cosmological timescales. For physical processes happening at shorter timescales, this variation will be negligible. 
We study here how the quasinormal modes of gravitational perturbations of a Schwarzschild-de Sitter (SdS) black hole will be affected by a variable effective cosmological constant $\Lambda$ as a function of the redshift $z$. The quasinormal mode frequencies will be a function only on the parameters of the spacetime \cite{Kokkotas,Nollert}. In the timescale for quasinormal mode oscillations,  $\Lambda$ will be essentially a constant, but black holes at different redshifts will have different quasinormal mode frequencies.

In Section \ref{sec:models} we present four different cosmological models and numerical results for each $\Lambda(z)$. The variation of the SdS quasinormal modes as a function of $\Lambda$ is shown in Section \ref{sec:qnms}. In Section \ref{sec:cosmo}  we propose a cosmological application of our results, calculate the resulting frequency shift for supermassive black holes and discuss the detectability of this effect. Finally, we give our final remarks in Section \ref{sec:final}.

\section{Cosmological models}
\label{sec:models}

We consider here four different cosmological models (models A-D), that have different underlying physical assumptions. From each one we can derive equations that describe the cosmological constant as a function of the redshift. We present numerical results for each model in \Fref{fig:lambda_z}. For all models we keep $z < 15$, which we consider as an upper bound for  black hole formation \cite{Volonteri2010}. 

We do not intend for this Section to be a comprehensive review on these models, so we will point the reader to the original papers in each subsection for more details. 

\subsection{Dimensional analysis}
\label{sec:model_A}

In \cite{Chen}, a model is proposed following some simple assumptions on quantum cosmology and a dimensional analysis. In this scenario, the effective cosmological constant is a function of the scale factor $a$ as $\Lambda \propto a^{-2}$ or, as an explicit function of the cosmological redshift $z$,
\begin{equation}
\Lambda(z) = \Lambda_0(1+z)^2\,,
\label{eq:lambda_A}
\end{equation}
where $\Lambda_0$ is the present value of the cosmological constant. This would lead to an early universe value of $\Lambda$ several orders of magnitude larger than the present $\Lambda_0$, which is a useful feature in light of the cosmological constant problem. This model is presented as ``model A'' in \Fref{fig:lambda_z}.

\subsection{Coupled quintessence}
\label{sec:model_B}

The coupled quintessence model proposed in \cite{Amendola2000} has a light scalar field $\phi$ (the quintessence) with an exponential potential $U(\phi) = Ae^{\sqrt{2/3}\kappa\mu\phi}$, linearly coupled to the ordinary matter, that is responsible for the cosmic acceleration as a dynamical cosmological constant. This model is written in terms of the variables
\begin{equation}
x_1 = \frac{\kappa}{H}\frac{\dot{\phi}}{\sqrt{6}}\,, \quad x_2 = \frac{\kappa a}{H}\sqrt{\frac{U}{3}}\,, \quad  
x_3 = \frac{\kappa a}{H}\sqrt{\frac{\rho_{r}}{3}}\,,
\label{eq:def_x1x2x3}
\end{equation}
where $H$ is the Hubble parameter and $\rho_r$ is the radiation density. By using the variables $x_1$, $x_2$ and $x_3$, the Klein-Gordon equation for the scalar field and the Friedman equation for the spacetime can be re-written as the following system
\begin{eqnarray}
x_1' &=&x_1\left( \frac{x_3'}{x_3} -1\right) -\mu x_2^2 + \beta(1-x_1^2-x_2^2-x_3^2)\,,\nonumber\\
x_2' &=& \mu x_1x_2 + x_2\left(2 + \frac{x_3'}{x_3}\right)\,,\nonumber\\
x_3' &=& -\frac{x_3}{2}(1-3x_1^2+3x_2^2-x_3^2)\,,
\label{eq:x1x2x3}
\end{eqnarray}
where the independent variable is $\alpha = \log a$, the prime denotes $d/d\alpha$ and $\beta$ is a dimensionless constant given by $\beta = \sqrt{3/2}C/\kappa$ ($C$ is the coupling constant between the scalar field and the ordinary matter). 

We remark here that the parameters $\beta$ and $\mu$ (related to the scalar field potential) are all that is needed to completely specify the model. However, the numerical solution found is strongly dependent on the initial values supplied for (\ref{eq:x1x2x3}). The initial values (at early times) required for cosmic solutions are near $(x_1,x_2,x_3) = (0,0,1)$ and the exact values used here were found by trial and error, until we approached the standard $\Lambda$CDM values for the present cosmological constant density parameter, $\Omega_{\Lambda_0} \approx 0.7$, and for the present matter density parameter, $\Omega_{m_0} \approx 0.3$. With these values, the model makes the transition from a radiation dominated universe, at early times, to a matter dominated universe until close to the present time, when we have a $\Lambda$-dominated stage.

With the numerical solution of the system (\ref{eq:x1x2x3}), we can then obtain the scalar field density parameter $\Omega_{\phi}$ as a function of $\alpha$,
\begin{equation}
\Omega_{\phi} = x_1^2+x_2^2 \equiv \frac{\rho_{\Lambda}}{\rho_c}\,,
\end{equation}
where $\rho_{\Lambda} = \Lambda/8\pi G$ and the critical density $\rho_c$ can be written as
\begin{equation}
\rho_c = \frac{\rho_{r_0}}{x_3^2}(1+z)^4\,,
\end{equation}
where $\rho_{r_0}$ is the present radiation density. Therefore, we can explicitly write
\begin{equation}
\Lambda(z) = \Lambda_0(1+z)^4\left(\frac{x_1^2+x_2^2}{x_3^2}\right)\left(\frac{x_{3_0}^2}{x_{1_0}^2+x_{2_0}^2}\right)\,, 
\label{eq:lambda_B}
\end{equation}
with $\Lambda_0 = 8\pi G \rho_{r_0}(x_{1_0}^2+x_{2_0}^2)/x_{3_0}^2$ and the $x_{i_0}$ are the present values of the $x_i$ at $z=0$. This model is presented as ``model B'' in \Fref{fig:lambda_z}.

Since there is no analytical solution for $\Lambda(z)$ in \Eref{eq:lambda_B}, we performed a 4th order polynomial fit for the numerical results, 
\begin{eqnarray}
\frac{\Lambda(z)}{\Lambda_0} &=& b_0 + b_1(1+z) + b_2(1+z)^2 + \nonumber \\
&+&b_3(1+z)^3 + b_4(1+z)^4\,,
\label{eq:lambda_B_fit}
\end{eqnarray}
with the coefficients given in the left side of \Tref{table:lambda_fit}.

\subsection{Interacting holographic dark energy model}
\label{sec:model_C}

In \cite{Elcio}, a holographic dark energy model with an interaction between dark energy and dark matter is studied. The future event horizon is chosen as the infrared cutoff, providing a time-dependent ratio of the matter energy density and the dark energy density. This model also allows the transition of the dark energy equation of state to phantom regimes, as proposed by \cite{Alam_etal2004}.

For this model we can write from the Friedman equation,
\begin{eqnarray}
 \label{eq:W_L_C}
 \frac{\Omega_{D}'}{\Omega_{D}^2} &=& (1-\Omega_{D})\left[\frac{1}{\Omega_{D}} + \frac{2}{c\sqrt{\Omega_{D}}} - \frac{3b^2}{\Omega_{D}(1-\Omega_{D})}\right]\,, \\
 \frac{H'}{H} &=& 3\left[\frac{1}{3} + \frac{2\sqrt{\Omega_{D}}}{3c} + \frac{b^2-1}{\Omega_{D}}\right]\frac{\Omega_{D}}{2}\,,
 \label{eq:H_C}
\end{eqnarray}
where $\Omega_D$ is the dark energy density parameter, $c$ is a constant and $b^2$ is the coupling constant for the interaction between the dark energy and the dark matter. After obtaining numerical solutions for $\Omega_{D}$ and $H$, we can finally write
\begin{equation}
 \Lambda \equiv \Lambda_0\frac{H^2}{H_0^2}\frac{\Omega_{D}}{\Omega_{D_0}}\,.
 \label{eq:lambda_C}
\end{equation}
where $\Lambda_0 = 3H_0^2\Omega_{D_0}$ and $H_0$ is the present value of the Hubble parameter $H(z)$. This model is presented as ``model C'' in \Fref{fig:lambda_z}.

Following the procedure we used in \ref{sec:model_B}, we did a polynomial fit of the numerical results obtained,
\begin{eqnarray}
\frac{\Lambda(z)}{\Lambda_0} &=& c_0 + c_1(1+z) + c_2(1+z)^2 + \nonumber\\
&+&c_3(1+z)^3 + c_4(1+z)^4\,,
\label{eq:lambda_C_fit}
\end{eqnarray}
with the coefficients given in the right side of \Tref{table:lambda_fit}.

\begin{table}[!htb]
\begin{center}
\begin{ruledtabular}
\begin{tabular}{ll|ll}
$b_0$ &  1.2161(58)            & $c_0$ & 0.6726(18)\\
$b_1$ &   - 0.2763(49)         & $c_1$ & 0.1936(15)\\
$b_2$ &   0.0800(12)           & $c_2$ & 0.11372(39)\\
$b_3$ &   0.03661(12)          & $c_3$ & - 0.002830(37)\\
$b_4$ &    $- 5.964(36) \times 10^{-4}$ & $c_4$ & $6.76(12) \times 10^{-5}$ \\
\end{tabular}
\end{ruledtabular}
\caption{Values for the $b_i$ and $c_i$ coefficients for the model B and model C fits given in \Eref{eq:lambda_B_fit} and \Eref{eq:lambda_C_fit}, respectively, and their standard deviations, calculated with a standard least squares fit using the numerical solutions presented in \Fref{fig:lambda_z}.}
\label{table:lambda_fit}
\end{center}
\end{table}

\subsection{Cosmological particle creation}
\label{sec:model_D}

Dark energy is represented by a dynamical cosmological constant in \cite{Saulo}. This model proposes that the spacetime expansion process can extract non-relativistic dark matter particles from the vacuum, at a constant rate $\Gamma$ given by
\begin{equation}
\Gamma = \frac{3}{2}(1-\Omega_{m_0})H_0\,,
\end{equation}
where $\Omega_{m_0}$ is the present matter density parameter. As a consequence, it can be shown that $\Lambda = 2\Gamma H$ and we can write it explicitly as 
\begin{equation}
\Lambda(z) =  \Lambda_0
[1-\Omega_{m_0} + \Omega_{m_0}(1+z)^{3/2}]\,.
\label{eq:lambda_D}
\end{equation}
where $\Lambda_0 = 3(1-\Omega_{m_0})H_0^2$. The concordance value obtained for $\Omega_{m_0}$ in this model with a joint analysis of the matter power spectrum, the position of the first peak of the CMB anisotropy spectrum and the Hubble diagram for type Ia supernovae is $\Omega_{m_0} \approx 0.45$ (higher than the usual $\Lambda$CDM
 value).

This is another example of a model with interaction in the dark sector. We can also cite as a motivation the fact that particle creation is expected to happen in curved spacetimes. The coincidence problem is considered to be alleviated in this context, because of the consequent slower decay of the $\Omega_m/\Omega_{\Lambda}$ ratio. This model is presented as ``model D'' in \Fref{fig:lambda_z}.

\subsection{Comparison between models A-D}

\begin{figure}[!htb]
\begin{center}
\includegraphics[angle=270,width=1\linewidth]{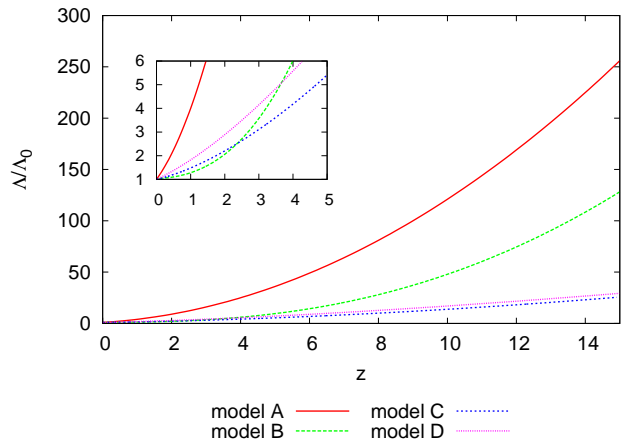}
\end{center}
\caption{$\Lambda/\Lambda_0$ as a function of $z$ for models A-D discussed in sections \ref{sec:model_A}-\ref{sec:model_D}. The analytical solutions for models A and D are given in (\ref{eq:lambda_A}) and (\ref{eq:lambda_D}). Model B is given by \Eref{eq:lambda_B}, where we used the numerical solution of the system (\ref{eq:x1x2x3}) with $\mu = 0.1$ and $\beta = 0.5$. Model C is given by \Eref{eq:lambda_C}, where we used the numerical solutions of \Eref{eq:W_L_C} and \Eref{eq:H_C} with $c = 1$ and $b^2 = 0.04$. }
\label{fig:lambda_z}
\end{figure}

The solutions obtained for $\Lambda(z)$ in the models A-D described in sections \ref{sec:model_A}-\ref{sec:model_D} are presented in \Fref{fig:lambda_z}. We can see that, in model A, $\Lambda$ grows much faster than the others, being larger by a factor 2 already at $z \approx 1$. For $0 < z \lessapprox 2.5$, the magnitude of $\Lambda$ follows model B $<$ model C $<$ model D. For $2.5 \lessapprox z \lessapprox 3.5$, we have  model C $<$ model B $<$ model D. Finally, for $z \gtrapprox 3.5$ we have model C $<$ model D $<$ model B.

Models C and D, even though motivated by different physical assumptions, provide very close results for the entire $z$ range considered. The difference between their values grows to  $\approx 15 \%$ at $z=15$. For $z \gtrapprox 3.5$, model B is intermediate between models C,D and model A, becoming larger than models C,D by a factor 2 at $z \approx 7$.

Based on these results, we can conclude that all models are distinguishable from each other solely by their $\Lambda(z)$ predictions. This could lead to a very direct way of comparing cosmological models, provided we have observational data on $\Lambda$ at different redshifts. In the next Section we will discuss how black hole oscillations could help us to obtain these results.

\section{Quasinormal modes for the Schwarzschild-de Sitter black hole}
\label{sec:qnms}

There is a vast literature on the quasinormal modes of stars and black holes, see \cite{Kokkotas, Nollert,Konoplya} for some very good reviews on the subject. The problem of finding the quasinormal modes of the SdS black hole has already been previously considered in the literature, see for instance \cite{Mellor1990,Otsuki1991,Moss2002,Zhidenko2003,Abdalla2004}, and it has gained more attention recently in the context of black hole thermodynamics and area quantization (see \cite{Choudhury2004,Skakala2012} and references therein). We review here only very briefly the most important points needed for our analysis.

The metric of a Schwarzschild-de Sitter spacetime with black hole mass $M$ and cosmological constant $\Lambda$ is described by
\begin{equation}
\label{eq:SdS}
ds^2 = -f(r)dt^2 + f(r)^{-1}dr^2 + r^2d\Omega^2\,, 
\end{equation}
where $d\Omega^2 = d\theta^2 + \sin^2\theta d\phi^2$ and $f(r)$ is given by
\begin{equation}
f(r) = 1-\frac{2M}{r} - \Lambda\frac{r^2}{3}\,.
\end{equation}

The radial component of a perturbation in this background satisfies a wave equation that can be put in the following form
\begin{equation}
\frac{\partial^2 \psi_{\ell}}{\partial t^2} + \left( -\frac{\partial^2}{\partial x^2} + V_{\ell}(r)\right)\psi_{\ell} = 0\,,
\label{psi}
\end{equation}
where $x$ is the tortoise coordinate given by $dr/dx = f(r)$ and the form of the potential $V_{\ell}(r)$ depends on the nature of the linear perturbations considered (see, for instance, the discussion in \cite{Nagar2005}).

The quasinormal modes are solutions to the perturbation equation (\ref{psi}) that satisfy the boundary conditions corresponding to purely outgoing waves at infinity and purely ingoing waves at the black hole horizon. These solutions are characterized by complex frequencies $\omega = \omega_r + i\omega_i$ (the quasinormal frequencies) that depend only on the parameters of the background spacetime, and not on the details of the initial perturbation.

\begin{figure}[!htb]
\begin{center}
\includegraphics[angle=270,width=1\linewidth]{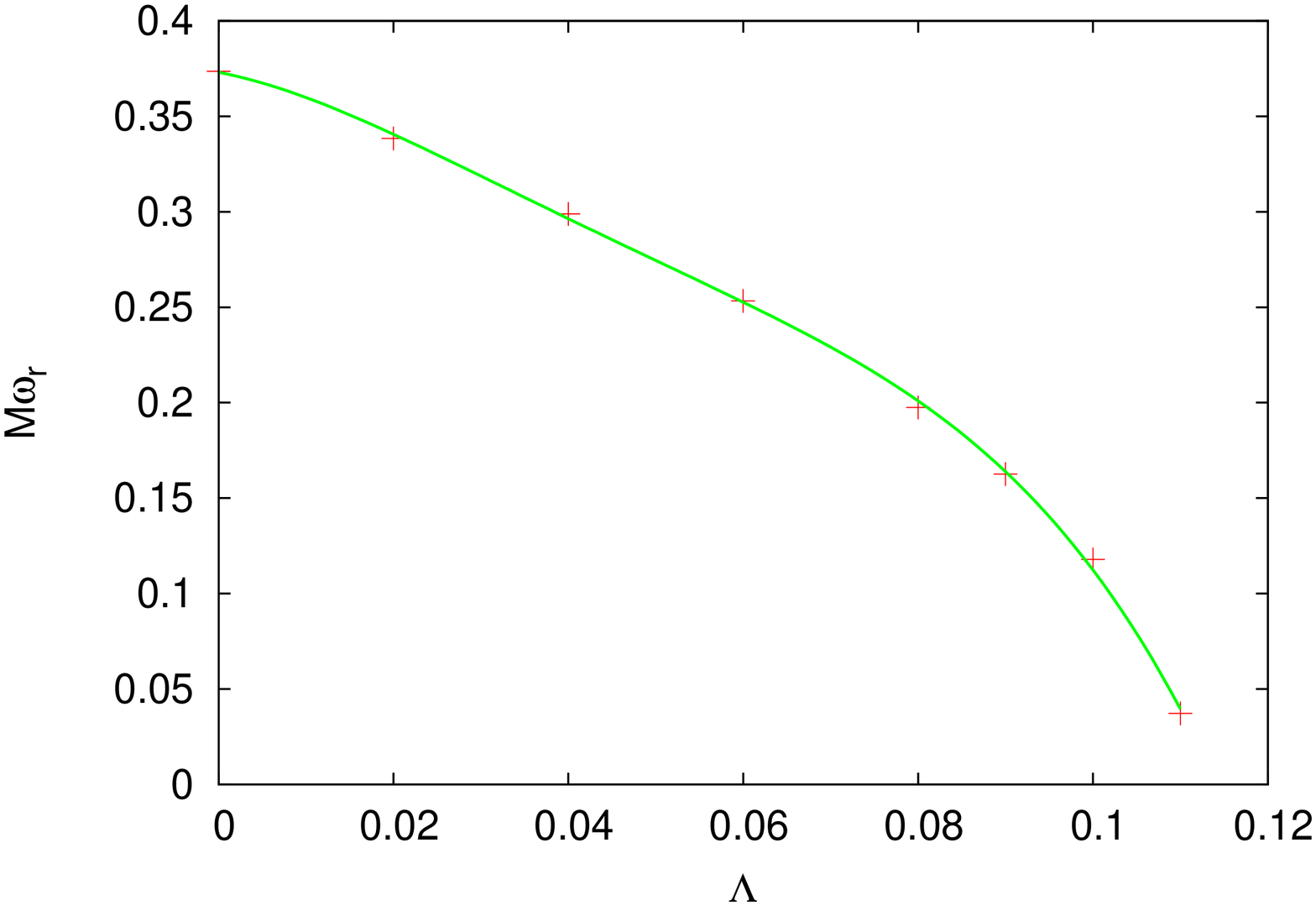}
\includegraphics[angle=270,width=1\linewidth]{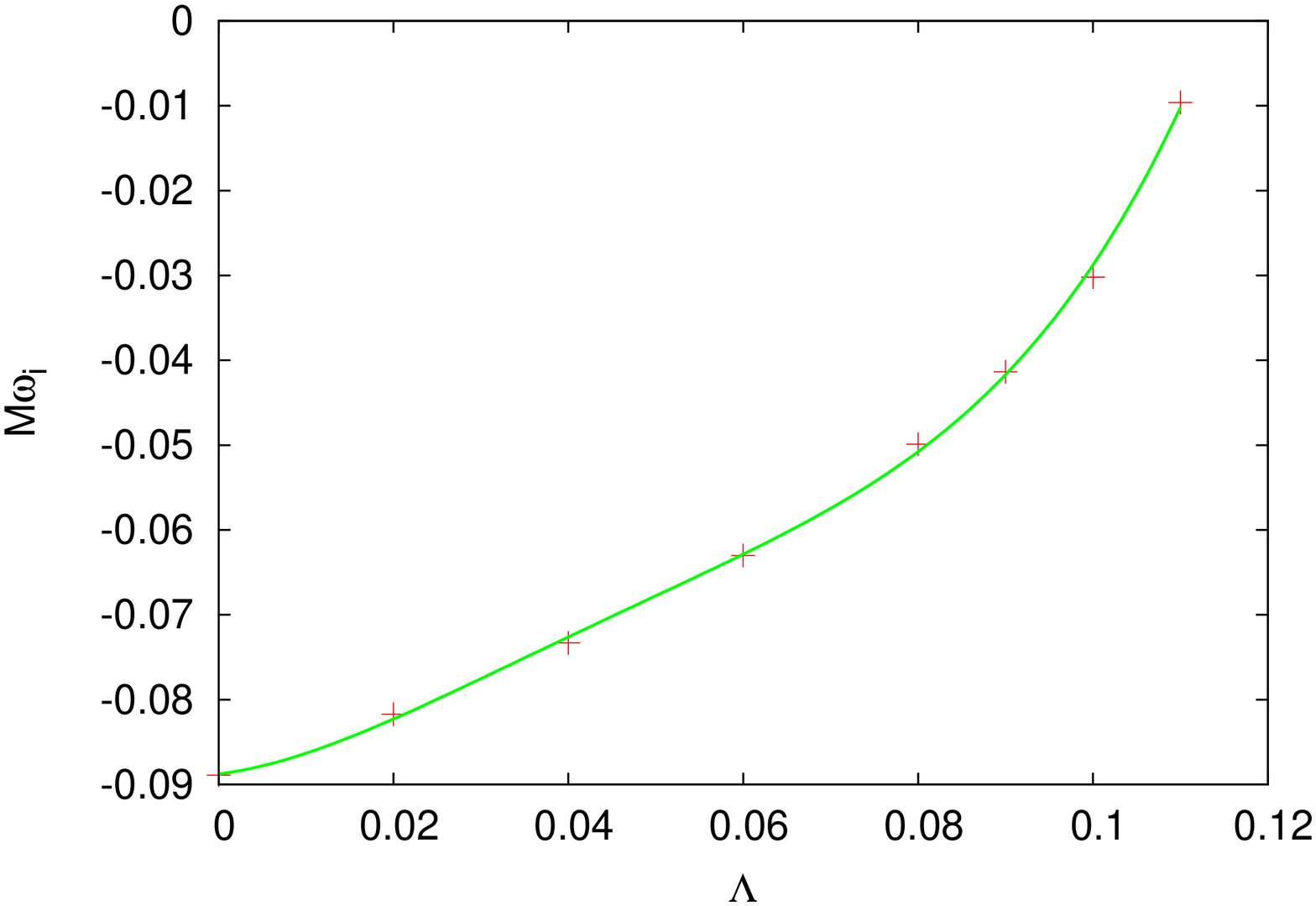}
\end{center}
\caption{Upper plot: $M\omega_r$ as a function of $\Lambda$ The data points are taken from \cite{Zhidenko2003} and the solid line represents the empirical fit (\ref{eq:omega_fit}), with the values for the coefficients given in Table \ref{table:omega_fit}. Lower plot: same as the upper plot, but for $M\omega_i$.}
\label{fig:omega_fit}
\end{figure}

For gravitational perturbations in the SdS spacetime, we can fit the dependence of the dimensionless quasinormal frequency $M\omega$ as a simple polynomial function of $\Lambda$:
\begin{equation}
 M\omega = a_0 + a_1\Lambda + a_2\Lambda^2 + a_3\Lambda^3 + a_4\Lambda^4\,,
\label{eq:omega_fit}
\end{equation}
where the complex coefficients $a_i$ are given in Table \ref{table:omega_fit}. Our empirical fits for $\omega_r$ and $\omega_i$ are presented in fig. \ref{fig:omega_fit}, together with numerical results taken from \cite{Zhidenko2003}. We chose a quartic fit based on the behavior of the data, as can be seen in the figure. We do not propose this empirical fit with a deeper analytical justification, but as a useful tool for our following analysis.

\begin{table}[!htb]
\begin{center}
\begin{ruledtabular}
\begin{tabular}{lll}
$a_i$  & $\textrm{Re}(a_i)$ & $\textrm{Im}(a_i)$\\
\hline
$a_0$ & 0.3730(45)   & -0.0888(12)\\
$a_1$ & -0.87(70)     & 0.13(18)\\
$a_2$ &  -53(28)       & 14.0(75) \\
$a_3$ & $9.0(40)\times 10^{2}$    & $-2.3(10) \times 10^{2}$\\
$a_4$ &  $-5.3(18) \times 10^{3}$ & $1.39(46) \times 10^{3}$      \\
\end{tabular}
\end{ruledtabular}
\caption{Values for the $a_i$ coefficients for the empirical fit given in eq. (\ref{eq:omega_fit}) and their standard deviations, calculated with a standard least squares fit using the data from \cite{Zhidenko2003} shown in fig. \ref{fig:omega_fit}, present in $c = G = 1$ units.}
\label{table:omega_fit}
\end{center}
\end{table}

Combining the empirical fits (\ref{eq:omega_fit}) with the solutions we obtained in Section \ref{sec:models} for $\Lambda(z)$, we can produce $\omega_r(\Lambda (z))$ and $\omega_i(\Lambda (z))$. 
These combined expressions show how the quasinormal mode frequencies would be affected by a variable cosmological constant: black holes at different redshifts will have different corrections from $\Lambda(z)$, that will depend on the particular cosmological model used. The necessary assumption implicit for this analysis is that the timescale for the evolution of $\Lambda$ is much longer than that of the black hole perturbations, which allows us to use the metric (\ref{eq:SdS}) with the variable $\Lambda(z)$.

\section{A cosmological application}
\label{sec:cosmo}

Since the actual value of $\Lambda_0$ is so small, we will keep now only the terms up to first order in $\Lambda$ in the absolute frequency difference caused by the time dependence of the cosmological constant, that can be written as
\begin{equation}
\Delta \omega_r \equiv (\omega_r-\omega_{0_r}) = \textrm{Re}(a_1)\frac{\Lambda_0}{M}\left(\frac{\Lambda(z)}{\Lambda_0}-1\right)\,,
\label{eq:Delta_omega}
\end{equation}
where $\omega_{0_r} = \omega_r(\Lambda_0)$. If we return now to physical units, we can assess exactly how small this effect is and the feasibility (or not) of using it as a real cosmological test.

For $\Lambda = 0$, the frequency $\bar{f}$ measured in Hz is
\begin{equation}
 \bar{f} = \frac{\textrm{Re}(a_0)}{2\pi}\left(\frac{c^3}{GM}\right) = 12.05 \left(\frac{M_{\odot}}{M}\right) \textrm{kHz}\,,
\end{equation}
and restoring the physical dimensions to $\Lambda$ and $\textrm{Re}(a_1)$ gives 
\begin{flushleft}
\begin{eqnarray}
 \bar{\Lambda} = \Lambda \left(\frac{c^2}{GM}\right)^2 = \Lambda \left(\frac{M_{\odot}}{M}\right)^2 4.58 \times 10^{-7}\textrm{m}^{-2},\\
 \textrm{Re}(\bar{a}_1) = 
  \textrm{Re}(a_1) \left(\frac{M}{M_{\odot}}\right)^2 2.18\times10^6 \textrm{m}^2,
 \end{eqnarray}
\end{flushleft}
(recall that $a_1\Lambda$ is dimensionless). If we take now the current value of the cosmological constant as approximately $\bar{\Lambda}_0 = 10^{-52} \textrm{m}^{-2}$, and $a_1 = -0.87$ from \Tref{table:omega_fit}, then we have $a_1\Lambda_0 = -1.9\times10^{-46}(M/M_{\odot})^2$ and we can finally rewrite \Eref{eq:Delta_omega} as
\begin{equation}
 \Delta\bar{f} = -6.14 \times 10^{-42}\left(\frac{M}{M_{\odot}}\right)\left(\frac{\Lambda(z)}{\Lambda_0}-1\right) \textrm{Hz}\,,
 \label{eq:Delta_f}
\end{equation}
which allows us to write the relative frequency shift 
\begin{equation}
 \frac{\Delta\bar{f}}{\bar{f}} = -5.10 \times 10^{-46}\left(\frac{M}{M_{\odot}}\right)^2\left(\frac{\Lambda(z)}{\Lambda_0}-1\right)\,.
 \label{eq:Delta_f/f}
\end{equation}
Equations (\ref{eq:Delta_f}) and (\ref{eq:Delta_f/f}) confirm the intuition that the effect of the cosmological constant on the quasinormal modes should be very small, but they go beyond this intuition and quantify the effect.\footnote{The analysis presented here can be reproduced for the damping time $\tau = 1/\omega_i$. The resulting very small variation in $\omega_i$ shows that the mode amplitude ratios will be almost identically the same as in the constant $\Lambda$ case.} We can see that it becomes larger at larger masses, but even in a case with a  $M=10^{10} M_{\odot}$ black hole and an enhancement factor of 100 from the evolution of $\Lambda$ there is no prospect of measuring $\Delta\bar{f}$. In this case, we would have the relative frequency shift $\Delta\bar{f}/\bar{f} \approx 10^{-24}$.

\section{Final remarks}
\label{sec:final}

We have presented here a comparison between four different cosmological models with variable cosmological constant. We presented analytical solutions for $\Lambda(z)$ when possible, in the case of models A and D (this result is new and was not presented in \cite{Saulo}) and numerical solutions with corresponding polynomial fits for models B and C, which are also novel results not provided in references \cite{Amendola2000} and \cite{Elcio}.

Our results show that all four models considered here can be distinguished from each other by their predicted evolutions, as can be seen in \Fref{fig:lambda_z}. Even for models C and D, which show the closest behavior, this difference amounts to $\approx 15\%$ at $z = 15$. For the other models, the difference can be as large as a factor $\approx 10$ at the same redshift.

We also studied the influence of the variable cosmological constant in the quasinormal modes of gravitational perturbations in a Schwarzschild-de Sitter black hole, and presented a numerical fit for $\omega(\Lambda)$. An increase in the value of $\Lambda$ will decrease both $\omega_r$ (and increase the oscillation period) and $|\omega_i|$ (and increase the damping time of the perturbation). 

The resulting effect in the quasinormal mode frequencies is extremely small. We quantified this effect and calculated it for a supermassive black hole as a cosmological application of our results. However, the detectability of the effect of the evolution of the cosmological constant in black hole quasinormal modes is very far from the current status of any gravitational wave detectors.

\acknowledgments
The authors wish to thank Elcio Abdalla, Saulo Carneiro, Cole Miller, Alberto Saa and Winfried Zimdahl for useful discussions and comments. This work was partially supported by the Brazilian agency CAPES (grant 2010/059582) and the Max Planck Society.

\end{document}